\documentclass[12pt, letterpaper]{article}

\usepackage{amsmath,amssymb}
\usepackage{fancyhdr}
\usepackage[top=.65in, bottom=.75in, left=1in, right=1in]{geometry}
\usepackage{graphicx}
\usepackage{hyperref}
\usepackage{float}
\usepackage{lscape}

\numberwithin{equation}{section}

\begin{document}


\setcounter{page}{0}
\date{}

\lhead{}\chead{}\rhead{\footnotesize{RUNHETC-2013-12\\SCIPP-13/8}}\lfoot{}\cfoot{}\rfoot{}

\title{\textbf{Two Point Pad\'{e} Approximants and Duality\vspace{0.4cm}}}

\author{Tom Banks$^{1,2}$ \and T.J. Torres$^{2}$ 
\vspace{0.7cm}\\
{\normalsize{$^1$NHETC and Department of Physics and Astronomy, Rutgers University}}\\
{\normalsize{Piscataway, NJ 08854-8019, USA}}\vspace{0.2cm}\\
{\normalsize{$^2$SCIPP and Department of Physics, University of California,}}\\
{\normalsize{Santa Cruz, CA 95064-1077, USA}}\vspace{0.2cm}\\
}

\maketitle
\thispagestyle{fancy} 

\begin{abstract}
\normalsize \noindent
We propose the use of two point Pad\'{e} approximants to find expressions valid uniformly in coupling constant for theories with both weak and strong coupling expansions. In particular, one can use these approximants in models with a strong/weak duality, when the symmetries do not determine exact expressions for some quantity.\end{abstract}


\newpage
\vspace{1cm}

\section{Introduction}

Many models in theoretical physics depend on a coupling parameter, $g$, and can be solved exactly in both the limits of weak and strong coupling.
Strong coupling expansions are less common in continuum quantum field theory, but they are possible in theories that have a strong/weak coupling duality.  Most of the work on such theories is devoted to exact results, but eventually we may want to resort to more artisanal methods to extract the physics of such models.  This is certainly the case in a variety of lattice models in condensed matter physics.

Recently, Sen\cite{sen} proposed an interpolating function for the mass of the $SO(32)$ spinor state in Type I/Heterotic string theory in $10$ dimensions.  His result captured both strong and weak coupling limits
and provided a smooth interpolation.  This motivated us to resurrect an ancient method for matching a pair of asymptotic series, whose convergence properties are well understood.  The basic idea is simple:  the weak and strong coupling asymptotics of a function $f(g)$ describing some physical property typically have the form of an exponential times a series of powers of the coupling parameter and its inverse.  In most cases there is only an essential singularity at {\it one} extreme of the coupling.  We can factor off this behavior and obtain a function $F(x)$ that has a simple power series in a variable $x = g^{\alpha}$ and $x^{-1}$ for ${\rm ln}\ x \rightarrow \pm\infty$.  Both asymptotic limits of $F(x)$ can then be captured by a rational function of $x$.

Two point Pad\'{e} approximants are rational functions whose coefficients are determined by matching some number of terms in both the $x$ and $x^{-1}$ series for $F(x)$.  Depending on how many terms we know in these series, we may be able to form a variety of approximants, and it is useful to look at all the possibilities in order to test the rate of convergence of the approximation. There are rather general theorems, which guarantee the convergence of near diagonal\footnote{These are approximants where the order of the numerator and denominator differ by a fixed integer $j$, and the sum of the orders goes to infinity. Often, one must throw out a sub-sequence of approximants, which has spurious poles in the domain of convergence. There are bounds on $j$, which depend on the nature of the singularities of the function being approximated.} Pad\'{e} approximants in rather general compact domains of the complex plane, including compact subsets of a strip parallel to the positive real axis\cite{ baker-book, scholar-pade}.  Note that while many of these theorems refer to functions regular at the origin, there are also strong convergence results for Pad\'{e} approximants to Stieltjes series \cite{baker}, and other functions with essential singularities at the origin.  Finally, note that the restriction to {\it compact } subsets of the positive real axis is removed if we consider the two point Pad\'{e} approximant, because convergence at $x \rightarrow \infty$ is guaranteed.

Experience shows that even when $F(x)$ has isolated singularities, the Pad\'{e} approximants try to match these singularities with poles or clusters of poles.  One can then try to probe the nature of the singularity.  For example, if there is a branch cut of the form  $(x - x_0)^{\beta}$, then $\frac{F^{\prime} (x)}{F(x)}$ will have a simple pole at $x_0$ and will be well approximated everywhere by the two point Pad\'{e} approximants.

In this paper, we will apply the Pad\'{e} technique to a number of examples, starting with Sen's spinor mass formula, and including the free energy, planar cusp anomalous dimension, and circular Wilson loop in $\mathcal{N}=4$ Super Yang-Mills theory.  We also indicate how to apply the technique to a variety of lattice spin and gauge models, but we do no explicit computations for these examples.  

\section{Review of the n-point Pad\'{e} Approximant Method}
Recall that the goal of any Pad\'{e} interpolation is to find a rational polynomial 
\begin{eqnarray*}
G_{N,M}(x) &=& P_N(x)/Q_M(x) \\
P_N(x) &=& \sum_{i=0}^{N}  p_i x^i \\
Q_M(x) &=& 1+ \sum_{i=1}^{M}  q_i x^i
\end{eqnarray*}

whose behavior matches that of a function $F(x)$ up to order $M+N+1$. That is to say
\begin{equation}
G_{N,M}(x) - F(x) = 0 + \mathcal{O}\left(x^{N+M+1} \right)
\end{equation}
The method for interpolating a function for a quantity whose series expansion behavior is known about a finite number of points is the following:

Let us construct a system of series expansions about some finite set of points, $x_i$ where $i \in \{1, \cdots, n\}$, and we will call these polynomials 
\begin{equation}
f_{i}(x) = \sum_{k=0}^{\infty} a_{i k} (x-x_i)^k
\end{equation}

We then require that the Pad\'{e} approximant match the series behavior up to order $j_i$ for each $f_i(x)$, and solve the resulting system of equations
\begin{equation*}
\begin{array}{rcl}
G_{N,M}(x) - f_1(x) & = & 0 + \mathcal{O}(x^{j_1})\\
G_{N,M}(x) - f_2(x) & = & 0 + \mathcal{O}(x^{j_2})\\
 &\vdots& \\
G_{N,M}(x) - f_n(x) & = & 0 + \mathcal{O}(x^{j_n})
\end{array}
\end{equation*}
with the constraint that $\sum_{i=1}^{n} j_i = N+M+1$ to obtain values for the coefficients $p_i$ and $q_i$. This procedure then gives a rational polynomial whose behavior matches $F(x)$ about each point $x_i$ up to order $j_i$.

\section{$SO(32)$ Spinor Masses in 10d Type I/Heterotic String Theory}
Here we show the use of two-point Pad\'{e} approximants when applied to the case of the masses of $SO(32)$ spinors, as in Sen's recent paper \cite{sen}.  The perturbative expression for the running of the $SO(32)$ spinor masses in Type I/Heterotic string theory admits a dual expansion in the limit of strong coupling. Thus we may use a Pad\'{e} approximant to match both the strong and weak coupling behaviors, thereby obtaining an expression for finite coupling that interpolates between the two dual expansions. In a revised version of his paper, published after our work was completed, Sen added a Pad\'{e} extrapolation of the spinor mass formula.  His technique differs from ours, in that he looks for rational functions of the original perturbative coupling, rather than finding a variable in which both weak and strong coupling series are ordinary power series. 

As Sen states, the asymptotic series expansions of the masses at strong and weak coupling are
\begin{eqnarray*}
F^W_2(g) &=& g^{1/4}\left(1 + K_w g^2+ \mathcal{O}(g^4) \right)\\
F^S_1(g) &=& g^{3/4}\left(1 + K_s g^{-1} + \mathcal{O}(g^{-2})\right)\\
K_w &\simeq& .23\\
K_s &\simeq& .351
\end{eqnarray*}
If we then let $g^{1/2}= y$ we can obtain convenient forms for the series expansions by factoring out a $y^{1/2}$ from $F^W_2$ and $F^S_1$ and then constructing the approximant
\begin{equation}
G_{N,M}(y)  = \frac{P_N(y) }{Q_M(y)} = \frac{ \sum_{n=0}^{N} a_n y^n}{1 + \sum_{m=1}^{M} b_m y^m}
\end{equation}
This gives the mass interpolating function as
\begin{equation}
M_{4,3}(y) = \sqrt{y} G_{4,3} = \frac{\sqrt{y} \left(y \left(1+K_w-K_s K_w\right)+y^4 K_w+y^3 K_w+y^2 K_w\right)}{1+ y \left(K_w-K_s K_w\right)+y^3 K_w+y^2 K_w}
\end{equation}
\begin{figure}[t]
\centering
\includegraphics[width = .8\textwidth]{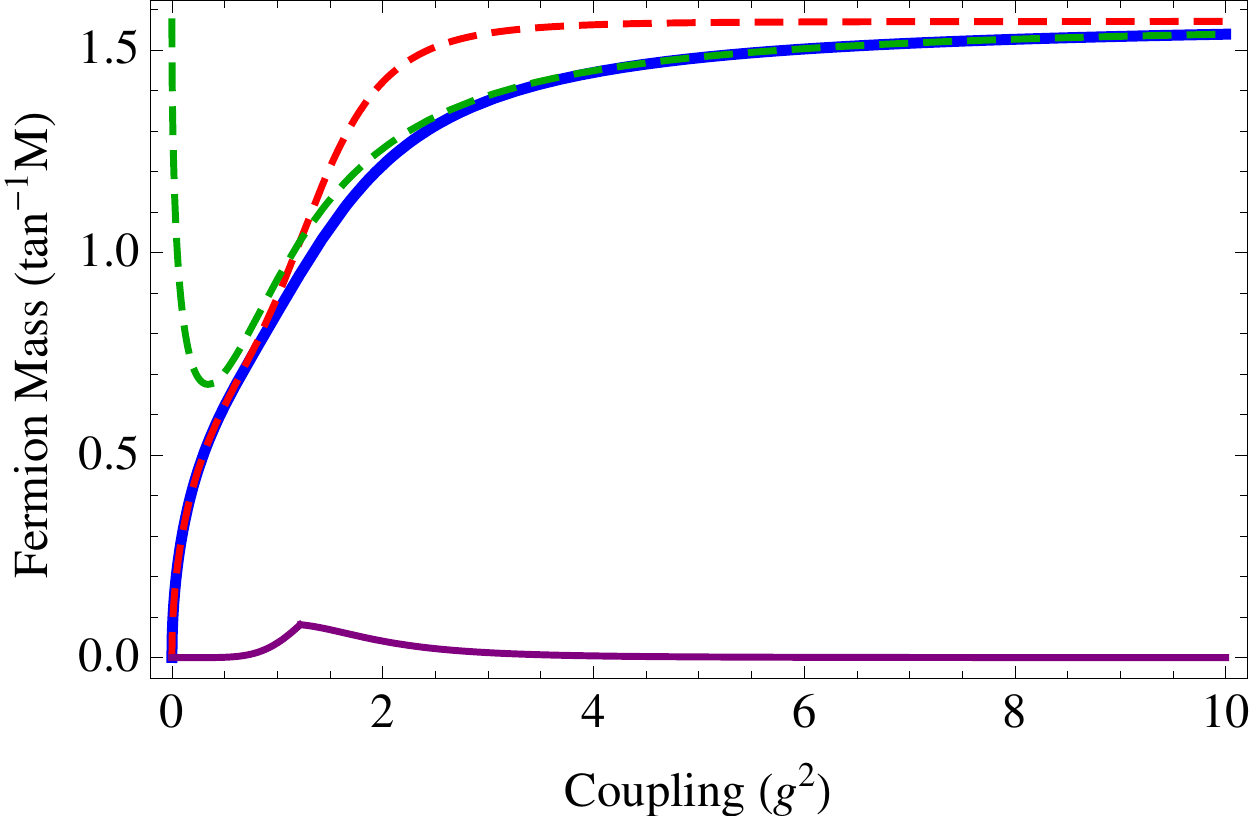}
\caption{Here we plot the minimal order, two-point Pad\'{e} approximant to the weak and strong coupling expansion of the spinor mass and compare it to the asymptotic expansions. The minimal order interpolating function, $M_{4,3}$, is in blue; the weak coupling expansion is in dashed red; the strong coupling expansion is in dashed green; and the minimal difference of the Pad\'{e} interpolation from the weak and strong coupling expansions is in purple. The maximal difference between the Pad\'{e} approximant and the asymptotic expansions is $(\Delta M)_{\text{max}} = 0.081$ at $g^2=1.28$, and the difference rapidly approaches zero to either side of the maximum. }
\label{SenPlot}
\end{figure}
The plot in Fig. \ref{SenPlot} shows the agreement between the two-point Pad\'{e} method and the asymptotic coupling expansions. One notes that the difference between the Pad\'{e} approximant and the series expansions is quite minimal throughout the entire coupling domain. 

\section{Thermal Free Energy in $\mathcal{N}=4$ Super Yang-Mills Theory}
\begin{figure}[t]
\centering
\includegraphics[width= .8 \textwidth]{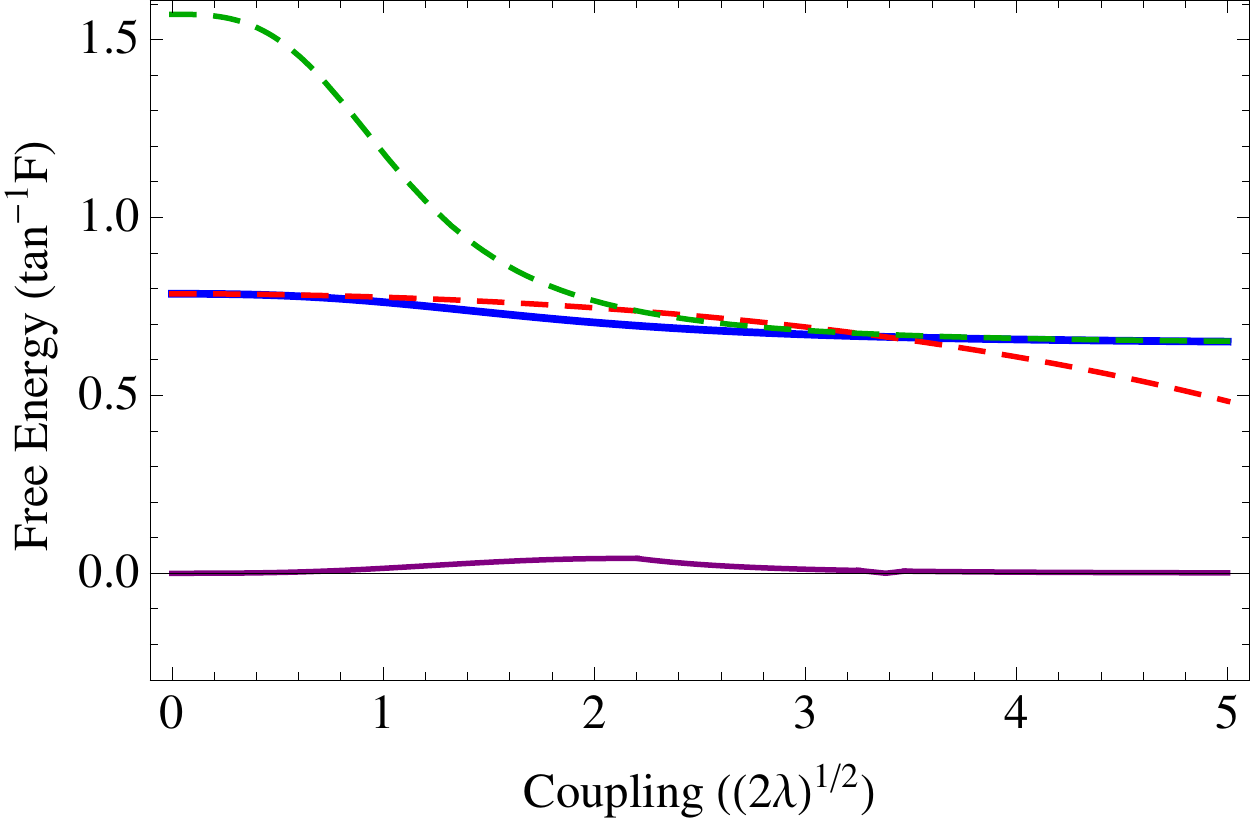}
\caption{Here we plot the minimal order, two-point Pad\'{e} approximant to the weak and strong coupling expansion of the thermal free energy in $\mathcal{N}=4$ SYM and compare it to the asymptotic expansions. The minimal order interpolating function, $G_{3,3}$, is in blue; the weak coupling expansion is in dashed red; the strong coupling expansion is in dashed green; and the minimal difference of the Pad\'{e} interpolation from the weak and strong coupling expansions is in purple. The maximal difference between the Pad\'{e} approximant and the asymptotic expansions is $(\Delta F)_{\text{max}} = 0.042$ at $x =2.14$, and again the difference rapidly approaches zero to either side of the maximum. }
\label{FreeEnergy}
\end{figure}

Previously calculated to lower order in the asymptotic expansions in \cite{kim}, Pad\'{e} approximants are also useful for interpolating thermal free energy between dual coupling regimes. In 3-dimensional $SU(N)$, $\mathcal{N}=4$ SYM the thermal free energy has dual expansions in the strong and weak coupling regime of the 'tHooft coupling, $\lambda$, given in \cite{taylor, gubser, burgess, rebhan} by
\begin{eqnarray}
F &=& \left(k_3 N^{2} V_3 T^{4} \right)F^{W,S}(x)\nonumber \\
F^W(x) &=& 1 -\frac{3}{4\pi^2} x^2 + \mathcal{O}(x^3)  \\
F^S(x) &=& \frac{3}{4} + \frac{45}{32} \zeta (3) x^{-3}  + \mathcal{O}(x^{-5})\\
x &=& \sqrt{2 \lambda} = \sqrt{2 g^2_{YM} N}\nonumber
\end{eqnarray}
where $T$ is the temperature and $V_3$ is the 3-dimensional volume, and $k_3$ is a factor of order unity\footnote{Originally we had taken the next highest order of the weak coupling expansion to be $\mathcal{O}(x^4)$, but as was pointed out by Anton Rebhan, the next order term has been calculated in \cite{nieto} and is $\mathcal{O}(x^3)$. The Pad\'{e} approximant matching with the higher order term included in the weak coupling expansion can be found in \cite{rebhan}.} . Setting the common factor in the expansions equal to one we find that the minimal Pad\'{e} approximant for this set of expansions is 
\begin{equation}
G_{3,3}(x) = \frac{1+\frac{9 x^2}{16 \pi ^2}+\frac{2 x^3}{15 \zeta (3)}}{1+\frac{3 x^2}{4 \pi ^2}+\frac{8 x^3}{45 \zeta
   (3)}}
\end{equation}
and its behavior, compared to the weak and strong coupling expansions, can be seen in Fig. \ref{FreeEnergy}. 

\section{The Large Angle Planar Cusp Anomalous Dimension in $\mathcal{N}=4$ Super Yang-Mills Theory}
In the limit of large angle, $\varphi \rightarrow \infty$, the planar cusp anomalous dimension is linear in angle with a coefficient that is the cusp anomalous dimension of a light-like Wilson loop, which depends only on the coupling \cite{correa}. 
\begin{equation}
\lim_{\varphi \rightarrow \infty} \Gamma_{\text{cusp}}(g,\varphi) \propto \varphi \, \Gamma^{\infty}_{\text{cusp}}(g)
\end{equation}

The null Wilson loop, $\Gamma^{\infty}_{\text{cusp}}$, at small $g$ is then calculated in \cite{korchemsky}. This scenario is dual to a string theoretic description, via gauge/string duality, of the classical energy of a large spin string in $AdS_5$. In this case, as discussed in \cite{frolov}, at large angular momentum, $S$, the classical energy scales as $E= S + \Gamma \ln S$. The divergent piece, $\Gamma$, is dual to the null Wilson loop. Thus, leaving off the common angular/spin factor, the two dual expansions are considered in \cite{frolov, beisert, kotikov, gubser2} and are given by
\begin{eqnarray}
\Gamma^W_{\text{cusp}} &=& 4 g^2 - \frac{4}{3} \pi^2 g^4 + \frac{44}{45} \pi^4 g^6 - 8\left(\frac{73}{630} \pi^6 + 4 \zeta(3)^2 \right) g^8 + \mathcal{O}\left(g^{10}\right) \\
\Gamma^S_{\text{cusp}} &=& 2 g - \frac{3 \ln{2}}{2 \pi} + \mathcal{O}\left(\frac{1}{g}\right)
\end{eqnarray}
These two equations are fit nicely by a Pad\'{e} approximant of order $G_{6,5}$ whose form is given numerically, for reasons of convenience of presentation, by
\begin{equation}
G_{6,5} = \frac{87.4384 x^6+97.1024 x^5+57.0406 x^4+11.1317 x^3+4 x^2}{43.7192 x^5+55.7857
   x^4+33.4311 x^3+17.55 x^2+2.78293 x+1}
\end{equation}

\begin{figure}[t]
\centering
\includegraphics[width= .8 \textwidth]{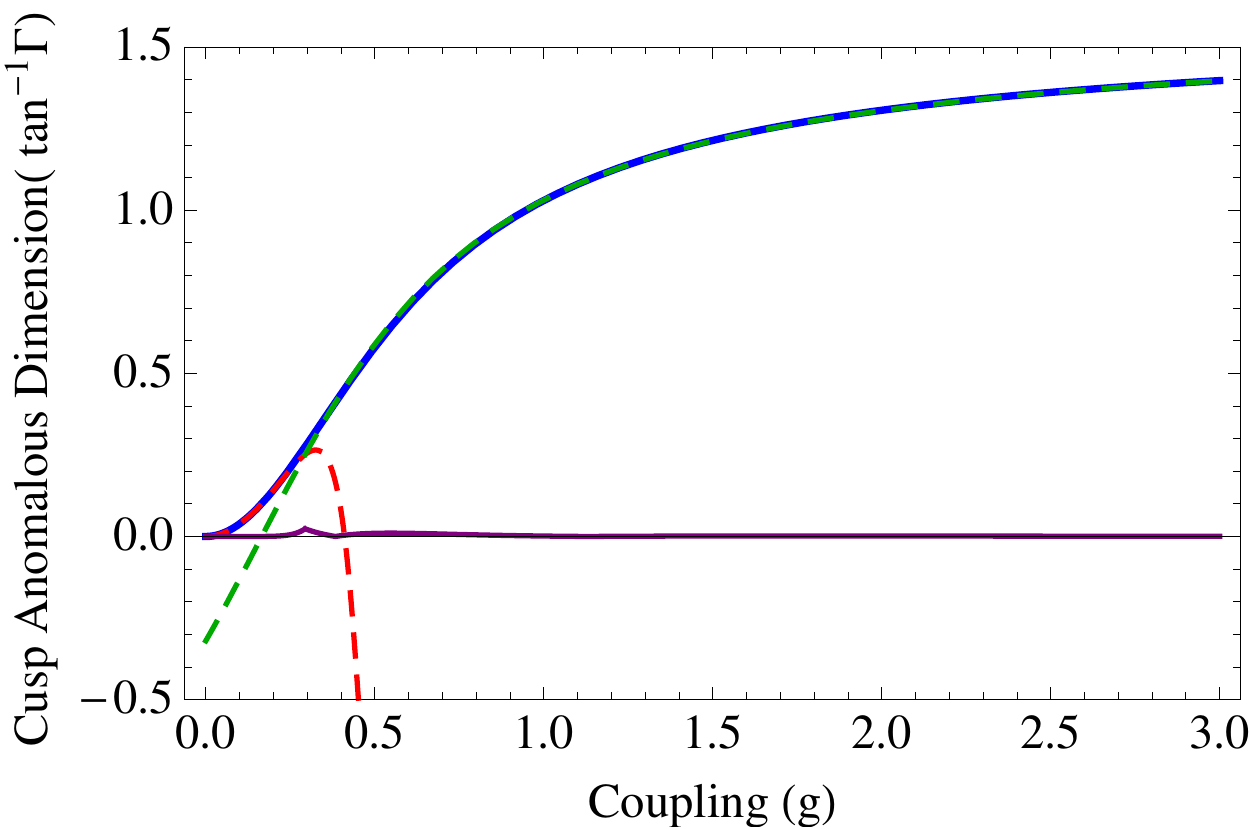}
\caption{Here we plot the minimal order, two-point Pad\'{e} approximant to the weak and strong coupling expansion of the cusp anomalous dimension in $\mathcal{N}=4$ SYM and compare it to the asymptotic expansions. The minimal order interpolating function, $G_{6,5}$, is in blue; the weak coupling expansion is in dashed red; the strong coupling expansion is in dashed green; and the minimal difference of the Pad\'{e} interpolation from the weak and strong coupling expansions is in purple. The maximal difference between the Pad\'{e} approximant and the asymptotic expansions is $(\Delta \Gamma)_{\text{max}} = 0.023$ at $g = .29$, and the difference rapidly approaches zero to either side of the maximum. }
\label{cusp}
\end{figure}

 Graphically, we can see in Fig. \ref{cusp} that the weak and strong behaviors are well interpolated by the approximant. 

\section{The $\mathcal{N}=4$ Super Yang-Mills Circular Wilson Loop }
Since there is an exactly calculable expression for the $SU(N)$ circular Wilson loop \cite{wilsonl_erickson, wilsonl_drukker}, it will be useful to check our Pad\'{e} Approximant with the exact theory. We can do this by calculating the asymptotic expansions in the strong and weak coupling limit, calculating the Pad\'{e} approximant from these expansions, then comparing to the exact expression. 

In the limit as $N$ goes to infinity for an $SU(N)$ gauge theory the exact circular Wilson loop is given by
\begin{equation}
\lim_{N \rightarrow \infty} \langle W_F \rangle = \frac{2}{\sqrt{\lambda}} I_1( \sqrt{\lambda})
\end{equation}
where $\lambda=g^2 N$ and $I_1$ is a modified Bessel function of the first kind. One can then calculate the asymptotic behavior of the Bessel (factoring out an exponential factor) to get
\begin{eqnarray}
\langle W_F \rangle^W &=& e^{\sqrt{\lambda} } \left( 1- \lambda^{1/2} + \frac{5 \lambda}{8} - \frac{7 \lambda^{3/2}}{24}+\frac{7 \lambda^2}{64}\right) \nonumber \\
\langle W_F \rangle^S &=& e^{\sqrt{\lambda} } \left( \sqrt{\frac{2}{\pi }} \left(\frac{1}{\lambda }\right)^{3/4}-\frac{3
   }{4 \sqrt{2 \pi }} \left(\frac{1}{\lambda }\right)^{5/4}\right)
\end{eqnarray}
where here we have factored out the exponential in order to match the polynomial behavior with a Pad\'{e} approximant and then multiply the end result by the exponential to determine the proper form for the Wilson loop. 

\begin{figure}[t]
\centering
\includegraphics[width= .8 \textwidth]{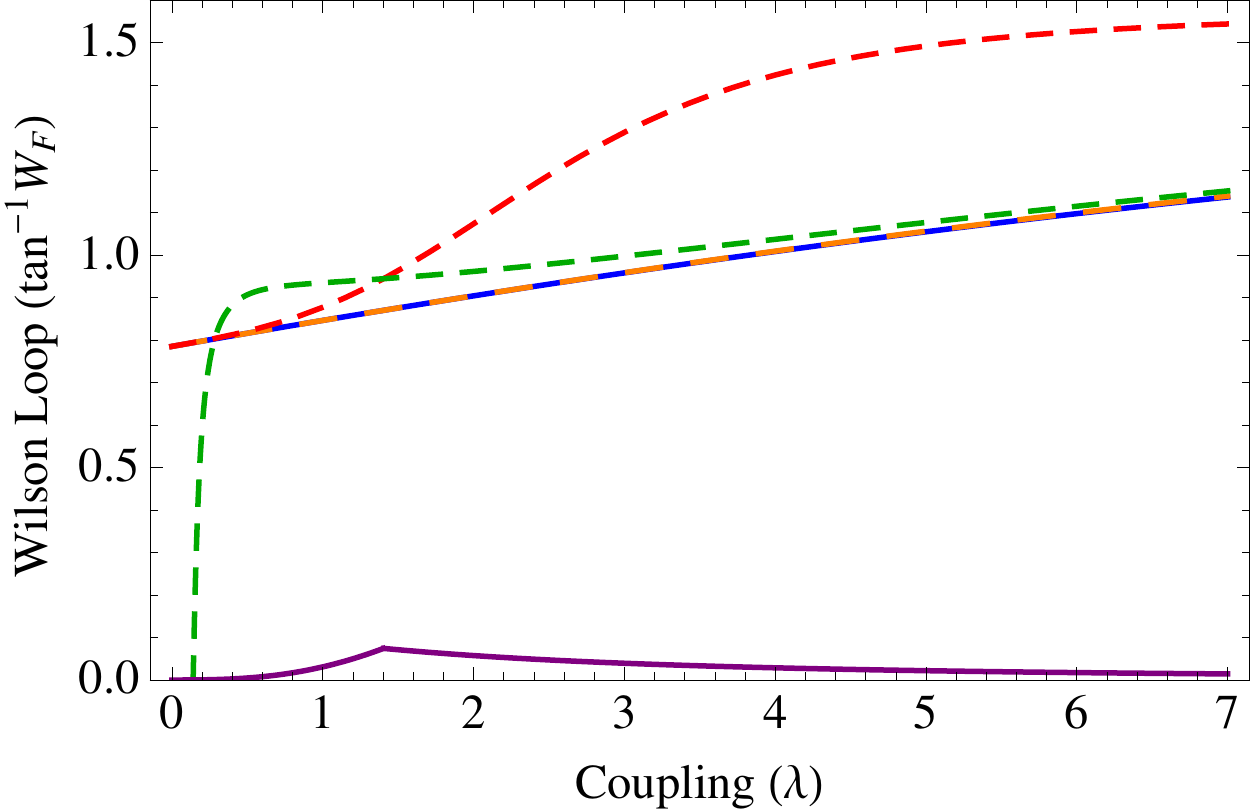}
\caption{Here we plot the minimal order, two-point Pad\'{e} approximant to the weak and strong coupling expansion of the circular Wilson loop in $\mathcal{N}=4$ SYM and compare it to both the asymptotic expansions and the exact expression for the Wilson loop. The minimal order interpolating function, $G_{5,7}$, is in blue; the weak coupling expansion is in dashed red; the strong coupling expansion is in dashed green; the exact result for the Wilson loop as $N \rightarrow \infty$ in dashed orange; and the minimal difference of the Pad\'{e} interpolation from the weak and strong coupling expansions is in purple. One can see that, while the Pad\'{e} approximant differs in the intermediate regime from the weak and strong coupling expansions, it matches the exact result to a very high degree of accuracy. }
\label{WilsonComp}
\end{figure}

\begin{figure}[H]
\centering
\includegraphics[width= .8 \textwidth]{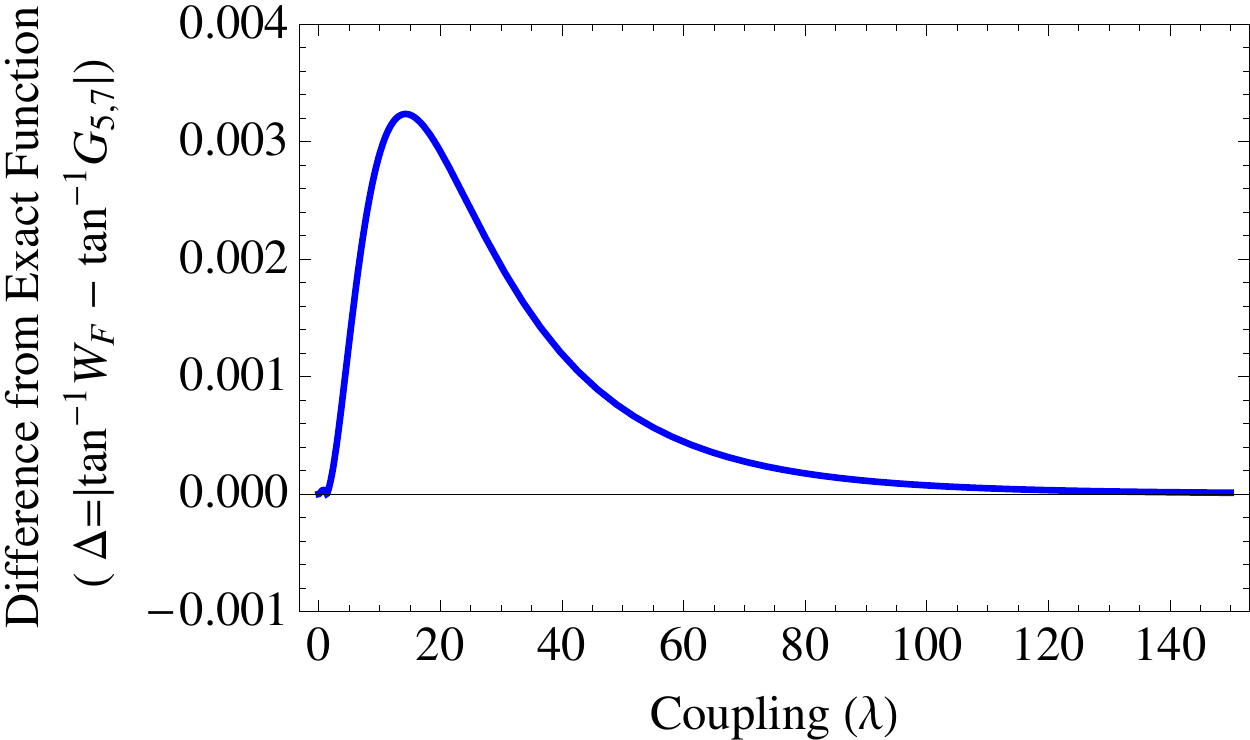}
\caption{This plot shows the difference between the exact, $N \rightarrow \infty$ limit of the circular Wilson loop in $\mathcal{N} = 4$ SYM and the Pad\'{e} approximant matching the asymptotic expansions of the exact function ($\Delta = | \tan^{-1}{W_F} - \tan^{-1}{G_{5,7}} |$). The relatively small difference between the exact and interpolated function shows that, in many cases, the approximant interpolation matches the exact formulation to a very high degree of accuracy throughout the entire domain of the coupling.}
\label{WilsonDiff}
\end{figure}

The two-point Pad\'{e} approximant is then given by Eq. \ref{circwilson} in the appendix and the matching behavior of the Pad\'{e} approximant when compared to the exact result is given in Fig. \ref{WilsonComp}. Furthermore, we provide a plot showing the quite minimal difference between the exact result and the approximant in  Fig. \ref{WilsonDiff}.

\section{Conclusion}

Here we have demonstrated the utility of two-point Pad\'{e} approximants to approximate the behavior of a variety of quantities in maximally supersymmetric YM theory with dual weak/strong coupling expansions. A rather general theorem guarantees convergence of the approximants around values of the coupling constant where the exact answer is analytic, and the two point approximants are guaranteed to converge in both the strong and weak coupling limits.  Furthermore, experience with Pad\'{e} extrapolation in condensed matter physics suggests that the Pad\'{e} method is useful for finding points of non-analyticity, which are caused by phase transitions.  

We have applied this method to several cases where duality allows us to make both a weak and strong coupling expansion. These cases include the $SO(32)$ spinor masses in Type I/Heterotic string theory, the thermal free energy of $\mathcal{N} = 4$  SYM theory, and the planar $\mathcal{N}=4$ SYM cusp anomalous dimension. We believe that the interpolations provided will well approximate the true values as a function of coupling. To that end we also show the method's application, as a test of its validity, to the case of the $\mathcal{N} = 4$ SYM circular Wilson loop. In this case an exact expression is known and we compared the interpolation to the exact form, showing very good agreement between the two. 

Two point Pad\'{e} approximants have a wide range of possible applications.  There are a number of other quantities in maximally SUSic YM theory to which they could be applied, including the anomalous dimensions studied in \cite{senrastalli}. In fact, the authors of this paper have pointed out to us that in one of its sections they propose a $\mathbb{Z}_2$ invariant re-summed perturbation series, which coincides with the maximal two point Pad\'{e} approximant.  Our approach would be more general, in that, with enough terms in the weak coupling series (which would generate a dual strong coupling series), we could generate a variety of two point Pad\'{e} approximants with the correct asymptotics and compare them to estimate the rate of convergence.  Given the paucity of terms in the extant series, this is not possible, and our proposal coincides with the $\mathbb{Z}_2$ invariant sum. 
	
Additionally, there are a large number of lattice models for which strong and weak coupling expansions are available.  $\mathbb{Z}_N$ spin and rank p gauge systems have a web of dualities in various dimensions, with a rank $p$ theory dual to a rank $d - p - 2$ theory.  The dual descriptions give strong and weak coupling expansions, which, for the Hamiltonian version, are simple expansions in powers of the basic coupling.  We also anticipate the possibility of using these methods to study versions of the Hubbard model that could be applicable to strongly correlated fermion systems.  We hope to return to these applications in future publications.

\section{Acknowledgements}
We would like to thank Carroll L. Wainwright for helpful discussions in the preparation of this work. Additionally, we'd like to acknowledge Santiago Peris for pointing out an important error regarding Pad\'{e} convergence, in the first version of this work. 

\newpage
\appendix
\begin{landscape}
\section{Long Approximant Forms}

\subsection{Circular Wilson Loop}
\begin{eqnarray}
&&G_{5,7} =  \label{circwilson} \\
   &&\frac{e^{\sqrt{\lambda }} \left(\frac{459 \sqrt{\frac{\pi }{2}} \lambda
   ^{3/4}}{1444+2256 \pi }+\frac{(247+96 \pi ) \lambda }{4332+6768 \pi }+\frac{(69 \pi
   -95) \lambda^{1/2} }{361+564 \pi }-\frac{\left(768 \pi ^{3/2}-2155 \sqrt{\pi
   }\right) \lambda^{1/4}}{\sqrt{2} (-1444-2256 \pi )}+1\right)}{-\frac{\left(247
   \sqrt{\pi }+96 \pi ^{3/2}\right) \lambda ^{7/4}}{\sqrt{2} (-4332-6768 \pi
   )}+\frac{459 \pi  \lambda ^{3/2}}{2888+4512 \pi }+\frac{3 \left(768 \pi ^{3/2}-931
   \sqrt{\pi }\right) \lambda ^{5/4}}{\sqrt{2} (11552+18048 \pi )}+\frac{4 \sqrt{2}
   \left(24 \pi ^{3/2}-53 \sqrt{\pi }\right) \lambda ^{3/4}}{361+564 \pi
   }+\frac{(1463+6924 \pi ) \lambda }{8664+13536 \pi }+\frac{(266+633 \pi )
   {\lambda^{1/2} }}{361+564 \pi }-\frac{\left(768 \pi ^{3/2}-2155 \sqrt{\pi }\right)
   {\lambda^{1/4} }}{\sqrt{2} (-1444-2256 \pi )}+1} \nonumber
\end{eqnarray}
   
   \end{landscape}

\bibliographystyle{./hunsrt}
\bibliography{pade}





\end{document}